\documentclass[twocolumn,preprintnumbers,amsmath,amssymb]{revtex4}
\usepackage{epsfig,graphicx}
\usepackage{dcolumn}
\usepackage{bm}
\begin{document}
\draft
\title{Molecular Dynamics Simulation of Vibrational Phase Relaxation along 
the Critical Isochore of Nitrogen : The Role of Local Density Fluctuations}
\author{Swapan Roychowdhury and Biman Bagchi\footnote[1] 
{For correspondence: bbagchi@sscu.iisc.ernet.in}}
\affiliation{Solid State and Structural chemistry Unit\\
Indian Institute of Science,\\
Bangalore - 560012, India}
\date{\today}
\begin{abstract}
 Vibrational dephasing of nitrogen molecule is known to show highly
interesting anomalies near its gas-liquid critical point. Here we 
present theoretical and computational studies of the Raman linewidth 
of nitrogen along the critical isochore. The linewidth is found to 
have a lambda shaped temperature dependence near the critical point. 
As observed in experimental studies, the calculated lineshape 
becomes Gaussian as the critical temperature ($T_{c}$) is approached. 
Both the present simulation and a mode coupling theory (MCT) analysis 
show that the slow decay of the enhanced density fluctuations near 
the critical point (CP), probed at the sub-picosecond time scales 
by vibrational frequency modulation, along with an enhanced 
vibration-rotation coupling, are the main causes of the observed 
anomalies.\\

    PACS numbers: 64.70.Fx, 82.53.Uv, 61.25.Em, 78.30.Cp
\end{abstract}
\maketitle
%\begin{multicols}{2}
 As the gas-liquid critical point of a fluid is approached, several  
dynamical properties of the system, such as the thermal conductivity, 
the bulk viscosity, the linewidth of the central Rayleigh spectrum,  show 
anomalous behaviour. This behaviour is collectively known as critical 
slowing down\cite{stanley,ma} and is physically explained by exploiting 
the divergence of the correlation length and using the dynamic mode coupling 
theory (MCT)\cite{fixman,ks}. In the recent years, several anomalies have 
also been observed in molecular relaxation processes near the critical point 
or in the supercritical fluid. One of them is the surprising augmentation 
of the solvent density around a solute\cite{mar}. Another 
interesting case is the  vibrational phase relaxation which shows more 
than one anomaly, yet to be understood.

    In their experiments, Clouter {\it et al.}\cite{clout} found that the 
isotropic Raman lineshape of a simple fluid like $N_{2}$ exhibits a 
remarkable additional non-rotational broadening near the gas-liquid
critical points ($\rho_{crit}, T_{crit}$). They measured   
the Raman spectra along the triple point to the critical point 
and the behaviour of the lineshape as the critical point is approached 
from above. Recently Musso {\it et al.}\cite{musso} measured the 
temperature dependence of the lineshape parameters (i.e shift, width 
and asymmetry) both along the coexistence and the critical isochore of 
liquid nitrogen and found that the temperature dependent linewidth 
($\Gamma$) is $\lambda$ shaped. The lineshape was found to undergo
a change from Lorentzian (away form $T_{c}$) to Gaussian (near $T_{c}$). 

  The observation of such anomalies has, till now, defied a convincing
explanation. Mukamel, Stern and Ronis \cite{muk} had earlier interpreted 
the rapid broadening of Raman lineshape as a manifestation of dynamical 
critical phenomena. This interpretation raises the following questions. 
As the microscopic time correlation function involved decays in less 
than a picosecond (actually in about 200 fs), why and how are the slow 
long wavelength density fluctuations important and relevant? 
The physics here is clearly different from the one involved in the critical 
slowing down at the long wavelengths. 

   In  a recent investigation, vibrational phase relaxation of the 
fundamental and the overtones of the N-N stretch in pure nitrogen was 
simulated by MD simulations, and the mode coupling theory 
(MCT)\cite{gay} was used to explain the simulation results. This study 
could reproduce the rapid rise in dephasing rate as the critical point 
is approached along the co-existence line, although it didn't include 
the vibrational coordinate(q) dependence of the inter-atomic potential 
and ignored the cross-term between the vibration-rotation coupling 
and force and their derivatives. Everitt and Skinner \cite{skin} studied 
the Raman lineshape of nitrogen in a systematic way by including of 
the bond length dependence of the dispersion and repulsive force 
parameters along the coexistence line of nitrogen. They also incorporated
the cross-correlation terms which were neglected earlier. Their results
for the lineshifts and linewidths along the gas-liquid coexistence of 
$N_{2}$ are in good agreement with experimental results. These theoretical
studies did {\em not} consider dephasing along the {\em critical isochore}. 
In addition, a convincing picture of anomalies did not emerge. The
results presented in this Letter provide first microscopic explanation of 
the above anomalies.

   The theories of the vibrational dephasing are all based on Kubo's 
stochastic theory of the lineshape\cite{kubo1}, extended to the study
of vibrational dephasing by Oxtoby\cite{ox}. The isotropic Raman 
lineshape, ${\it I(\omega)}$, is the Fourier transform of the normal 
coordinate time correlation function, $C_{Q}(t)$\cite{ox}, 
\begin{equation}
I(\omega)\;=\;\int_{0}^{\infty}\exp(i\omega t)\left[<Q(t)Q(0)>\right].
\end{equation}
   A cumulant expansion of Eq.(1) followed by truncation after second 
order gives the following well-known expression of $C_{Q}(t)$\cite{kubo2},
\begin{eqnarray}
<Q(t)Q(0)>\;=\;Re\;\exp(i\omega_{0}t+i<\Delta\omega>t)\nonumber \\
\times\;\exp\left[-\int_{0}^{t}dt^{\prime}(t-t^{\prime})<\Delta\omega(
t^{\prime})\Delta\omega(0)>\right].
\end{eqnarray}
  The frequency modulation time correlation function, $C_{\omega}(t)=<\Delta\omega(t)\Delta
\omega(0)>$, derives contributions from the atom-atom(AA), resonance(Rs),
vibration-rotation(VR) coupling, and also the cross-terms\cite{ox}. 
We calculated the linewidth, the lineshape and the dephasing time of $N_{2}$ 
for different thermodynamical state points of nitrogen, both along the 
coexistence line and the critical isochore, using Eqs. 1 and 2.

  The Hamiltonian of homonuclear diatomic molecules can be written as
\begin{equation}
{\rm H}(\vec q)\;=\;{\rm H}_{\it v} + {\rm T}(\vec q) + {\rm U}(\vec q),
\end{equation}
 where ${\rm H}_{\it v}$ is the vibrational Hamiltonian, ${\rm T}(\vec q)$
is the total translational and rotational kinetic energy, ${\rm U}(\vec q)$ 
is the inter-molecular potential energy, and ${\vec q}$ is the collection 
of vibration coordinates $q_{\it i}$.
 The inter-molecular potential energy
is sum of the following site-site (${\it v}_{ij}$) between two molecules {\it i} 
and {\it j}$\;$\cite{skin}. 
   The total intermolecular potential energy is taken to be
\begin{eqnarray}
{\rm U(\vec q)}&=&\frac{1}{2}\sum_{i\ne j}\sum_{\alpha\beta}{\it v}_{\it ij}
\left(\epsilon_{ij},\sigma_{ij},r_{i\alpha j\beta}\right)\nonumber\\
&=&\frac{1}{2}\sum_{i\ne j}\sum_{\alpha,\beta}^{\it 1,2} 4
{\epsilon}_{i\alpha j\beta}\left[\left(\frac{{\it \sigma}_{i\alpha j\beta}}
{{\rm r}_{i\alpha j\beta}}\right)^{12}-\left(\frac{{\sigma}_{i\alpha j\beta}}
{{\rm r}_{i\alpha j\beta}}\right)^6\right],
\end{eqnarray}
 where  $r_{i\alpha j\beta}=|{\vec {r}}_{j\beta}(q_{j}) - {\vec r}_{i\alpha}
(q_{i})|$ and ${\vec r}_{j\beta}(q_{j})={\vec r}_{j\beta}(0) + {\hat r}_{j
\beta}q_{j}/2$. ${\vec r}_{j\beta}(q_{j})$ is the position of the nucleus 
of atom $\beta$ in molecule $j$, and $\hat {\rm r_{j}}_\beta$ is the unit 
vector oriented from the center-of-mass of molecule j to the $\beta$th atom.
  The vibrational coordinate dependence of Lennard-Jones parameters are given 
by, $\epsilon_{i\alpha j\beta}=\sqrt{\epsilon_{i\alpha}\epsilon_{j\beta}}
=\epsilon(1+\gamma q_{i}+\gamma q_{j} + 2\gamma^{2}q_{i}q_{j})$ and
$\sigma_{i\alpha j\beta}=\left(\frac{\sigma_{\alpha\alpha}+\sigma_{\beta\beta}
}{2}\right)=\sigma(1+\delta q_{i}+\delta q_{j})$. 
We use the linear expansion 
coefficients $\gamma$ and $\delta$ as determined by Everitt and Skinner \cite{skin,param}.

    Microcanonical (NVE) MD simulations\cite{allen} were carried along the 
coexistence line and along the critical isochore using the leap-frog 
algorithm for different thermodynamical state points of nitrogen. A system 
of 256 diatomic particles was enclosed in a cubic box, and periodic boundary 
conditions were used. In simulation, the system was allowed to equilibrate for 
100 000 time steps with $\Delta t=\;0.0002\tau$, where $\tau$ [$=\sqrt{(m
\sigma^{2}/\epsilon)}$, m being the mass of the molecule ] is found to be equal 
to 3.147 ps. The averages were calculated over 400 000 MD steps. The 
thermodynamic state of the system can be expressed in terms of the reduced 
density of $\rho^{*}=\rho\sigma^{3}$ and a reduced temperature of $T^{*}= 
k_{B}T/\epsilon$. The density of the system has been expressed in terms of 
number of molecules per unit volume times $\sigma^{3}$ and the temperature 
is in units of $\epsilon/k_{B}$. Where $\sigma$ and $k_{B}$ is the 
Lennard-Jones diameter of the molecule and is the Boltzmann constant,
respectively. Limited number of simulations have been done with N = 512
 molecules. We found no significant difference in the linewidth at the
larger system.

\vspace*{0.1cm}
\begin{figure}[htb]
\epsfig{file=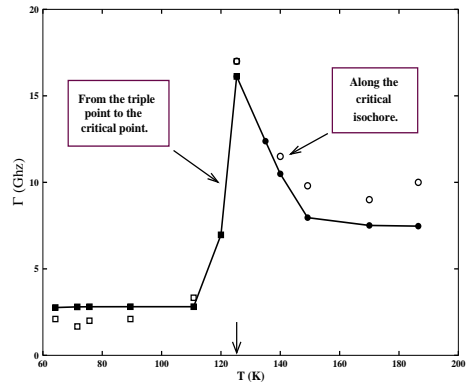,height=6cm,width=5.0cm,angle=-90}
\caption{The simulated lambda ($\lambda$) shaped linewidth ($\Gamma$) of 
nitrogen along the coexistence line (solid square)  and the critical 
isochore (solid circle). The experimental results for linewidth along
coexistence line (open square) and  the critical isochore (open circle)   
reported by Clouter and Kiefte and by Chesnoy in Ref. \cite{clout} are 
also shown. The CP is indicated by an arrow on the abscissa.}
\label{fig1}
\end{figure}

    Figure $1$ shows both the simulated temperature dependent dephasing rate of 
nitrogen and also the experimental one. 
The solid squares are the linewidth for different state points along 
the coexistence curve of nitrogen and  the solid circles are  the linewidth
along the critical isochore. The interesting feature in the figure is the 
lambda-shaped linewidth when the values for two different regions of nitrogen 
are presented together. This figure is  similar to the one observed 
experimentally (see figure $4$ of Ref.\cite{musso}). It is interesting to note 
the sharp rise in the dephasing rate as the CP is approached. There are noticeable
difference in the high temperature region along the critical isochore. 	 

  To understand the origin of this critical behaviour, we carefully analyzed
each of the six terms\cite{six} (three auto correlations and three cross-terms 
between density, vibration-rotation coupling and resonance), which are  
responsible for the modulation of the vibrational frequency for fundamental 
transition. Two terms are found to dominate near the CP and 
these are the density and the vibration-rotation coupling. The temperature 
dependence of these two terms are shown in figure $2$ where the integrand of
Eq.2, $X_{ij}(t)=\int_{0}^{t}dt^{\prime}(t-t^{\prime})C_{\omega}^{ij}
(t^{\prime})$, is plotted against time for seven state points along the 
critical isochore. Where i, j represents the density(den), vibration-rotation 
coupling(VR) and resonance(Rs) terms respectively. Note the sharp {\it rise} in 
the value of the integrand as the critical temperature is approached, and the 
{\it fall} when it is crossed. We have found that both these contributions at 
the CP are distinct compared to the other state points. Thus, the 
rise and fall of dephasing rate arises partly from the rise and fall in the 
density and the vibration-rotation terms.

\vspace*{0.44cm}
\begin{figure}[htb]
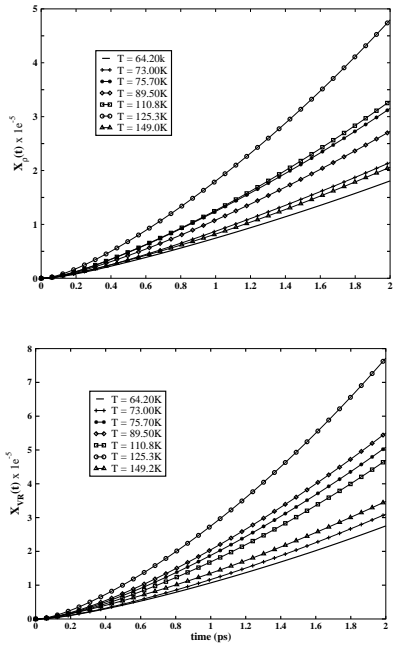

\epsfysize=4cm
\epsffile{fig2a.eps}\\
\vspace*{0.5cm}
\epsfysize=4cm
\epsffile{fig2b.eps}
\label{fig2}
\caption{The time dependence of the relative contributions of the (a) 
density-density, $X_{den}(t)$, and (b) VR coupling, $X_{VR}(t)$, terms at 
different temperatures along the critical isochore.}
\end{figure}

  A crossover from Lorentzian-like to Gaussian lineshape can happen when the 
usually large separation in the time scales of decay of $C_{\omega}(t)$ 
and $C_{Q}(t)$ \cite{musso,ox} ceases to exist and the two time correlation 
functions begin to overlap. Indeed, the computed lineshape becomes Gaussian 
near the from CP. Note that the frequency modulation time correlation function 
decays fully in about 200 femtosecond.

  What could be the reason for this dramatic crossover behaviour?    
The increase in density fluctuations\cite{muk,bin} near the CP 
increases the magnitude of the mean square frequency fluctuation 
$<\Delta\omega^{2}_{i}>$. This leads itself to an increase in the rate 
of decay of the normal coordinate time correlation function, ($C_{Q}(t)$). 
In addition, the frequency modulation time correlation function slows down 
somewhat. As the decay of these two time correlation functions become comparable, 
the lineshape goes over to the Gaussian form. This a novel effect and different 
from that commonly encountered.

  If one considers only the number density as the relevant slow variable
in dephasing, then mode coupling theory analysis (MCT) gives the following expression 
for the density dependence frequency modulation time correlation function\cite{sarika},
\begin{eqnarray}
\left<\Delta\omega_{\rho}(0)\Delta\omega_{\rho}(t)\right> = \frac{k_{B}T}{6\pi^{2}\hbar^{2}\rho}
\int_{0}^{\infty} k^{2} dk F_{s}(k,t)C^{2}(k)F(k,t),
\end{eqnarray}
where $C(k)$ is the Fourier Transform of two particle direct correlation function.
Near CP, the main contribution is derived from the long wavelength (that is small 
$k$) region, where self-intermediate scattering function, $F_{s}(k, t)= 
e^{-D_{s}k^{2}t}$ and intermediate scattering function, $F(k,t)
= S(k)e^{-D_{T}k^{2}t}$, $D_{s}$ is the self-diffusion coefficient, 
where $S(k)$ is the static structure factor, and $D_{T}$ is the thermal 
diffusivity. Thus, $\left<\Delta\omega_{\rho}(0)\Delta\omega_{\rho}(t)\right> 
\simeq S(k \rightarrow 0)e^{-(D_{s}+D_{T})k^{2}t}$. Near  
CP, $S(k\rightarrow 0)$ becomes very large ( as compressibility diverges 
at $T = T_{c}$ ), leading towards a Gaussian behaviour for lineshape. $D_{T}$
also undergoes a slowdown near $T_{c}$. This complex dependence may lead to 
a Levy distribution from time dependence of $\left<Q(0)Q(t)\right>$ as 
discussed earlier by Mukamel et al.\cite{muk} However, a limitation of the 
above analysis is the absence of the VR term which contributes significantly 
and may mask some of the critical effects. At high temperature, the latter 
dominates over the density term. A complete Lorentzian behaviour is predicted 
only in the low temperature liquid phase. Interestingly, the predicted 
divergence of $<\Delta\omega^{2}(0)>$ very close to $T_{c}$ enhances the
rate of dephasing and this shifts the decay of $<Q(t)Q(0)>$ to short times, 
giving rise to the Gaussian behaviour. We shall return to this point later. 
We have used MCT to demonstrate that the large enhancement of  Vibration-Rotation 
coupling near the gas-liquid critical point arises from the non-Gaussian behavior 
of density  fluctuation and this enters through a non-zero value of the triplet 
direct correlation function.

  To further explore the origin of these anomalous critical temperature 
effects, we have investigated for the presence of dynamical heterogeneities 
in the fluid at three temperatures near the CP, by calculating 
the well-known non-Gaussian parameter $\alpha(t)$ defined as\cite{ara}, 
$\alpha(t) = \left(\frac{3}{5}\right)\frac{<r^{4}(t)>}{<r^{2}(t)>^{2}} - 1$
 where $ <\Delta r(t)^{2}>$ is the mean squared displacement and $ <\Delta 
r(t)^{4}>$ the mean quartic displacement of the center of mass of nitrogen 
molecule. It can only approach zero (and hence Gaussian behaviour) for times 
exceeding the time scale required for individual particles to sample their 
complete kinetic environments. As can be seen from figure $3$, the function 
$\alpha$(t) is large near CP at times 0.5 - 5 ps, indicating the 
presence of long lived heterogeneities near $T_{c}$. 

\begin{figure}[htb]
\centerline{
\epsfxsize=6.0cm
\epsffile{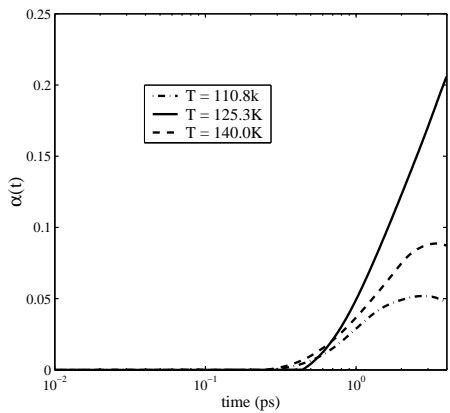}}
\caption{The non-Gaussian parameter $\alpha(t)$ is plotted against time (t) 
at three different temperatures along the critical isochore.}
\label{fig3}
\end{figure}

 The contribution of the density term reflects the combined effects of 
collisions, repulsive and attractive interactions on the friction on the 
bond. This term naturally grows as the CP is approached. The rise in the 
contribution of the vibration-rotation term has a different origin. This 
increase is due to the slowing down of the rotational time correlation 
function near the CP\cite{zewail}. As we approach the CP, the number of 
molecules having slow decay of rotational time correlation function increases. 
While one can describe the average effect of density fluctuation through 
the divergence of the static structure factor $S(k)$ at small wavenumbers 
(as discussed above), this approach does not capture the full scenario. 
This is because the decay of the frequency modulation time correlation 
function ($C_{\omega}(t)$) occurs in the femtosecond time scale. At that 
time scale, the density fluctuation is nearly static. Thus, vibrational 
dephasing provides snap shots of the large density fluctuations present 
near the critical temperature. Even more interesting is the origin of the 
ultrafast decay of $C_{\omega}(t)$--this is partly due to the cancellation 
which arises from the cross-terms of VR coupling with the density and the 
resonance terms. These terms have sign opposite to the pure terms but have 
comparable magnitude, leading to further enhancement in the rate of decay 
of $C_{\omega}(t)$. The reason for such large negative cross-correlation 
can be understood in terms of the inhomogeneity. Note that critical anomaly 
is nearly absent in $CF_{4}$, $CH_{4}$ and $CO_{2}$\cite{refB}. While cubic 
symmetry in the the former two precludes VR coupling, $CO_{2}$ may be too 
heavy for VR to be important.

  It is indeed surprising that our simulations could capture 
 many of the novel features observed in experiments, including the lambda shaped 
temperature dependence of the dephasing rate and the cross-over from the
Lorentzian to the Gaussian form. Large density fluctuations near $T_{c}$
shifts the dynamics probed to sub-picosecond times. This combined with VR coupling,
give rise to the observed anomalies.

 We thank P. Jose, A. Mukherjee,
and R. K. Murarka for helpful discussions. SR acknowledges the CSIR (India)
for financial support. This work is supported in part by grants from DAE 
and CSIR, India.

%\end{multicols}

\begin{references}
\bibitem{param} Parameters for $N_{2}$; L-J parameters  are $r_{o}/\AA$ = 
1.094 ,$\epsilon/kK$ = 37.3, and $\sigma /\AA$ = 3.31, m/amu = 28.0; 
Spectroscopic constant, $\omega_{o}/cm^{-1}$ = 2358.57 and Expansion 
coefficients are $\gamma /\AA^{-1}$ = 0.62, $\delta /\AA^{-1}$ = -0.063
\end{references}
\end{document}